\documentclass[aps,pre,amsmath,amssymb,aps,10pt,twocolumn,floatfix,superscriptaddress]{revtex4-1}
\usepackage{color,soul}
\usepackage{graphicx}
\usepackage{dcolumn}
\usepackage{bm}
\usepackage[usenames,dvipsnames]{xcolor}
\usepackage[colorlinks=true,citecolor=Blue,linkcolor=RubineRed,urlcolor=Blue]{hyperref}

\begin{document}

\title{Problem hardness of diluted Ising models: Population Annealing versus Simulated Annealing}

\author{Fernando Mart\'{i}nez-Garc\'{i}a}
\email{f.martinez@iff.csic.es}
\affiliation{%
Instituto de Física Fundamental IFF-CSIC, Calle Serrano 113b, Madrid 28006, Spain
}

\author{Diego Porras}
\affiliation{%
Instituto de Física Fundamental IFF-CSIC, Calle Serrano 113b, Madrid 28006, Spain
}

\begin{abstract}
    Population annealing is a variant of the simulated annealing algorithm that improves the quality of the thermalization process in systems with rough free-energy landscapes by introducing a resampling process. 
    We consider the diluted Sherrington-Kirkpatrick Ising model using population annealing to study its efficiency in finding solutions to combinatorial optimization problems. 
    From this study, we find an easy-hard-easy transition in the model hardness as the problem instances become more diluted, and associate this behaviour to the clusterization and connectivity of the underlying Erd\H{o}s-R{\'e}nyi graphs. We calculate the efficiency of obtaining minimum energy configurations and find that population annealing outperforms simulated annealing for the cases close to this hardness peak while reaching similar efficiencies in the easy limits. Finally, it is known that population annealing can be used to define an adaptive inverse temperature annealing schedule. We compare this adaptive method to a linear schedule and find that the adaptive method achieves improved efficiencies while being robust against final temperature miscalibrations. 
\end{abstract}

\maketitle

\section{Introduction}
The field of combinatorial optimization aims to find a solution that minimizes an objective function from a set of finite candidate solutions. Although this set of potential solutions is finite, it is often so large that searching for the optimal solution through brute force is not practical. Many tasks of interest can be framed as combinatorial optimization problems, making this field of research highly relevant in industry - with examples such as logistics~\cite{Cerny1985,johnson1990local} and finance \cite{Mansini1999,CramaSchyns2003,Rubio2024physA} - 
and science, where a paradigmatic example is the study of low-temperature phases and ground state configurations of statistical physics models such as Ising Hamiltonians. Amongst the latter, some of the most challenging models are those describing spin-glasses~\cite{edwards1975theory, mezard1987spin}.
In particular, studying the ground states of three and higher-dimensional Ising spin glasses is known to be an NP-hard combinatorial optimization problem~\cite{barahona1982computational, jauma2024exploring}.

Due to the difficulty and importance of finding solutions to complex statistical physics models, creating and improving algorithms for finding ground states has been the subject of considerable effort. The most popular approach is based on Monte Carlo algorithms, such as simulated annealing~\cite{kirkpatrick1983optimization}, parallel tempering~\cite{swendsen1986replica}, or population annealing~\cite{hukushima2003population, machta2010population} to produce approximate samples from the Boltzmann distribution. These methods have demonstrated their utility for finding ground states and studying the thermal properties of these systems. 
Furthermore, Monte Carlo methods have proved very useful in providing heuristics for combinatorial problems beyond the field of statistical physics \cite{kirkpatrick1983optimization,Cerny1985,CramaSchyns2003,Rubio2024physA}.
However, the application of these methods to general combinatorial optimization problems faces significant differences with respect to the problems for which they were originally developed. General optimization problems in logistics or finance typically lack the regular connectivities between variables that are found in spin-lattice models. Furthermore, objective functions can take complex mathematical forms that complicate estimating typical energy scales and choosing hyperparameters such as annealing temperatures \cite{rubiogarcía2022arXiv}. Population annealing is an auspicious approach to addressing these limitations. As we explain in more detail below, this algorithm provides us with an improved method for sampling Boltzmann distributions, together with an adaptive temperature schedule that facilitates its application to a diverse range of objective functions. 
Given the potential of population annealing for combinatorial optimization, it is relevant to gain insights into what advantages it offers depending on the characteristics of the targeted optimization problems.

The Population Annealing (PA) algorithm is a modified version of Simulated Annealing (SA). 
Both algorithms attempt to generate samples distributed according to a target thermal distribution at a given temperature. This is done by following the Metropolis-Hastings rule~\cite{robert1999monte} at intermediate temperatures following an annealing temperature schedule that slowly brings the system to the target temperature. However, the PA algorithm differs from SA by introducing a resampling step between samples (also called \emph{replicas}) at each temperature step. This step helps the system to stay closer to the correct thermal distribution by relocating samples that would otherwise be trapped at a local minimum. The effect of this additional step has been studied for spin-glasses in a 3D Edwards-Anderson model with Gaussian disorder~\cite{wang2015comparing} and demonstrated an improved efficiency of PA over SA for finding ground states for this family of problems, and a similar efficiency as parallel tempering. However, these results can be complemented by studying the problem hardness and the relative efficiency improvement of PA over SA for other problem families.

In this work, we study the relative behaviour of PA compared to SA when finding ground state configurations of Ising models with spin interactions defined by random graphs. The study of random graphs is related to a variety of real-world problems, with examples such as internet topology~\cite{albert1999diameter, faloutsos1999power}, social networks~\cite{newman2002random, robins2007recent}, and spread of diseases through populations~\cite{newman2002spread}. In this work, we consider the case of the diluted Sherrington-Kirkpatrick Ising model~\cite{boettcher2020ground} and study the relative efficiency between the SA and PA algorithms as a function of the connectivity of the models under consideration (parameterized by a value $p$), from sparse graphs up to all-to-all graphs. 
By considering this family of problems and using PA to obtain the values of the \textit{entropic family size}~\cite{wang2015population} and the \textit{mean square family size}~\cite{amey2018analysis} -- which are related to problem hardness -- we observe an easy-hard-easy transition in problem hardness as the problems become more diluted, which we associate to the clusterization and connectivity of the underlying Erd\H{o}s-R{\'e}nyi graphs~\cite{erdos1959random, erdos1960evolution, bollobas1998random, frieze2015introduction}. After studying this behaviour, we compare the efficiency of SA and PA to find the ground state of this type of problem. These results show that PA has a considerable efficiency advantage over SA for the problem instances generated with a connectivity value around the hardness peak, which vanishes as the connectivity moves away from this value. This is consistent with a physical picture in which PA is advantageous in hard combinatorial optimization problems with intermediate connectivities, where individual samples are likely to get stuck in local minima. Moreover, we expand this study by considering a version of PA that employs an adaptive temperature schedule~\cite{barash2017gpu, amey2018analysis} and demonstrate that this schedule offers a further efficiency improvement as compared to a linear inverse temperature schedule while also being more robust against final temperature miscalibrations.

This work is organized as follows: In
Section~\ref{sec:methods} we introduce the diluted Sherrington-Kirkpatrick Ising model, which will be the focus of our study. We will also provide a brief explanation of the simulated annealing and population annealing algorithms, and introduce the quantities that we will use to characterize the problem hardness and the relative efficiency between these algorithms. In Section~\ref{sec:problem_hardness} we investigate the problem hardness by using population annealing simulations and study an easy-hard-easy transition as a function of the density of interactions between spins. We continue by exploring how this problem hardness transition affects the relative efficiency between simulated annealing and population annealing in Section~\ref{sec:PAvsSA}, while also benchmarking the efficiency of the adaptive temperature schedule offered by population annealing. Finally, in Section~\ref{sec:conclusions} we present our conclusion to this study.

\section{Model and Methods}
\label{sec:methods}

\subsection{Diluted SK Model}
The Sherrington-Kirkpartrick (SK)~\cite{sherrington1975solvable} Ising spin-glass Hamiltonian is defined as:
\begin{equation}
\label{eq:ising_hamiltonian}
    \mathcal{H}=-\frac{1}{2}\sum^N_{i,j} J_{ij}s_i s_j,
\end{equation}
where $s_i=\pm 1$ represents the value of the $i$-th spin, of a system of $N$ spins. The matrix $J$ is a symmetric matrix where the entries $J_{ij}$ are sampled from a Gaussian distribution with mean zero and variance of $1/N$. In this work, we study a variation of this model that consists of a spin glass with spin interactions defined by an Erd\H{o}s-R{\'e}nyi graph, also known as diluted SK model~\cite{boettcher2020ground}. This model is similar to the SK model but with the particularity of setting random bonds equal to zero (both $J_{ij}$ and its symmetric $J_{ji}$) with a probability $1-p$ to achieve the connectivity expected from an Erd\H{o}s-R{\'e}nyi graph. To ensure a similar behaviour for different values of $p$, we set the variance of the entries $J_{ij}$ to $1/Np$, so that the ground state energy is similar for each value of $p$ considered. To avoid confusion, we clarify that two common variants of Erd\H{o}s-R{\'e}nyi graphs are frequently studied: the $G(N,p)$ model, which we have just described, and the $G(N,D)$ model, in which each node has exactly $D$ edges. The diluted SK model that we consider is defined over $G(N,p)$ Erd\H{o}s-R{\'e}nyi graphs.

\begin{figure}[t]
\centering
\includegraphics[width=0.8\columnwidth]{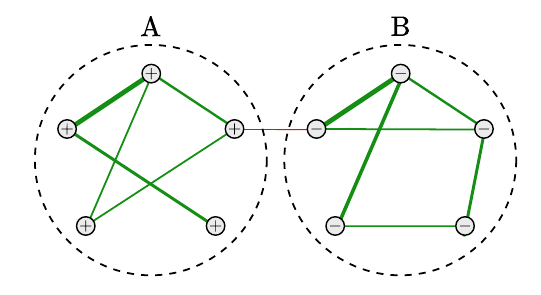}
\caption{Example of a problematic spin configuration for finding the minimum energy configuration using single-spin flip algorithms. For this case, we considered all $J_{ij}\geq 0$, with bigger widths representing more positive values of $J_{ij}$, and the colors indicate whether the interaction contributes with negative energy (green) or not (red). While clusters A and B are in a local minimum energy configuration, one of these clusters should flip all of its spins to reach the global minimum. However, if the temperature is low, it is unlikely to achieve this by performing single-spin flips.}
\label{fig:diluted_example}
\end{figure}

The minimization of models such as this one can be problematic when using local search algorithms, like the ones considered in this work which depend on single-spin flip updates. This is due to the appearance of local minima that are separated from the global minimum by high energy barriers~\cite{svenson2001freezing, katzgraber2008spin}. An example of this behaviour is shown in Fig.~\ref{fig:diluted_example}, in which the appearance of weakly interacting clusters can considerably complicate the combinatorial optimization process. In this work, we argue that population annealing, thanks to its resampling step, can help alleviate this type of complication. Therefore, considering the diluted SK model will allow us to study and compare the performance of the simulated annealing and population annealing algorithms, which we introduce in the following, as a function of the connectivity of the problem instances.

\subsection{Simulated Annealing}

Simulated Annealing (SA) is a Monte Carlo algorithm that starts with a set of $R$ \textit{replicas} $\mathbf{s}^{(r)}$, $r=1,..., R$. In our case, these replicas are spin configurations, which have an energy associated with a given Hamiltonian such as the one introduced in Eq.~\eqref{eq:ising_hamiltonian}. When the spin values of these replicas are initialized randomly, they can be considered as samples from a Boltzmann distribution at a temperature $T_0\rightarrow \infty$ or inverse temperature $\beta_0=1/T_0\rightarrow 0$. Given a replica $\mathbf{s}$ with energy $E$, it is possible to propose a single-spin flip change that results in a new configuration $\mathbf{s}'$ with energy $E'$. Using the Metropolis rule~\cite{robert1999monte}:
\begin{equation}
\label{eq:MH_rule}
    P_\text{accept}(\mathbf{s}'|\mathbf{s}) = \begin{cases}
        1 \quad &\text{if} \quad E' \leq E \\
        e^{-\beta_1(E'-E)} &\text{if}  \quad E' > E ,
    \end{cases}
\end{equation}
it is possible to accept or reject changes of this type so that, after proposing enough changes, the initial replicas obtained from 
a thermal distribution at inverse temperature $\beta_0$ are transformed into approximate samples corresponding to a thermal distribution at $\beta_1>\beta_0$. In our case, the single-spin flips are proposed by performing a total of $N_S$ ordered sweeps over all the spins. The replicas obtained, that approximate samples obtained at inverse temperature $\beta_1$, can then be used to obtain samples at a higher value of $\beta$. This defines an iterative process that anneals the replicas to increasing inverse temperature values $\beta_t$, with $t=1,...,N_T$, until reaching a target final temperature. These steps constitute the SA algorithm.

In SA, the $R$ replicas are independent of each other. Thus, the probability of successfully finding the ground state using $R$ replicas, $g(R)$, can be obtained from the probability of finding it with a single copy, $g(1)$, as:
\begin{equation}
\label{eq:gR}
    g(R) = 1 - \left[1-g(1)\right]^R.
\end{equation}
Using this expression one can estimate the so-called Time-To-Solution (TTS)~\cite{weigel2007genetic, ronnow2014defining, wang2015comparing, perera2020computational}, which is the time needed to reach a specific success rate (we will consider $99\%$). The TTS is then defined as:
\begin{equation}
    \text{TTS}_{0.99}=t_{SF} N_S N_T N \frac{\log(1-0.99)}{\log(1-g(1))},
\end{equation}
where $t_{SF}$ is the time required to perform a sweep. In the following, we simply set $t_{SF} = 1$ and define the TTS as the number of computational steps.

\subsection{Population Annealing}
\label{sec:population_annealing}

Population Annealing (PA)~\cite{hukushima2003population, machta2010population} is a sequential Monte Carlo algorithm~\cite{doucet2001introduction} that works similarly to SA but introduces a resampling step when changing the temperature of the system from $\beta_t$ to $\beta_{t+1}$. This resampling is performed between the replicas by associating a probability proportional to the relative Boltzmann weights between each temperature to each replica:
\begin{equation}
    \tau_i = \frac{e^{-(\beta_{t+1} - \beta_t) E_i}}{Q(\beta_t, \beta_{t+1})},
\end{equation}
with the normalization factor
\begin{equation}
    Q(\beta_t, \beta_{t+1}) = \sum_{i=1}^R e^{-(\beta_{t+1} - \beta_t) E_i}.
\end{equation}
There are different ways of resampling using these probabilities. In this work, we implemented the so-called \textit{systematic resampling} approach~\cite{gessert2023resampling, martinez2024near}. This resampling scheme can be visualized by positioning each replica in a line with an associated length equal to its corresponding $\tau_i$ value, thus dividing the $[0,1)$ interval into $R$ segments. The resampling is then implemented by generating a random number $U_0\in [0,1/R)$ and selecting $R$ positions within the $[0,1)$ interval given by the values $U_k=U_0+k/R$ with $k=0,..., R-1$. Each value of $k$ will be associated with the replica corresponding to the segment where $U_k$ lays, leading to that replica being resampled. This method keeps the number of replicas constant and requires only one randomly generated number for each resampling step. The resampling step helps the algorithm avoid local minima, for which the SA algorithm has no mechanism to escape and would otherwise be trapped, resulting in a reduced number of effective replicas exploring the configuration space. While the resampling process generates copies of configurations, thus introducing correlations, this is alleviated after enough Monte Carlo steps, which decorrelates them through single-spin flip changes. Finally, we note that the time required for the resampling steps is negligible compared to the time needed for the spin-flip part of the algorithm, making the efficiency comparisons between SA and PA straightforward.

An additional advantage of PA is that one can use the statistical information of the samples at each temperature to implement an adaptive inverse temperature schedule. Examples of adaptive schedules propose temperature changes that resample a given fraction (\textit{culling fraction}, $\varepsilon$) of the population at each step~\cite{amey2018analysis}, or that enforce a given overlap between energy histograms at consecutive temperature steps~\cite{barash2017gpu}. In this work, we will consider the culling fraction approach. While finding the inverse temperature change given a fixed culling fraction can be done by using algorithms such as bisection search, we will use the following simple rule derived from considering that at step $t$ the replicas energies are given by a Gaussian distribution with standard deviation $\sigma_{E,t}$:
\begin{equation}
    \Delta\beta_t \approx \frac{\epsilon \sqrt{2\pi}}{\sigma_{E,t}}.
\end{equation}
We note that we use this approximated adaptive schedule for simplicity in both the implementation and the conceptual design of the algorithm. However, recent progress has been made in the development of more advanced adaptive schedules, and results that improve over the ones shown in this work might be achieved by considering non-approximated schedules based on Ref.~\cite{amey2018analysis} or Ref.~\cite{barash2017gpu}, or based on the recent results found in Ref.~\cite{barzegar2024optimal}. In this work, we will consider a linear inverse temperature schedule when not considering the adaptive one.

Finally, while it is possible to study the problem hardness by using SA to find the TTS, this approach becomes computationally expensive since it first requires finding the ground state of each problem, and then using SA with enough replicas $R$ such as to obtain a good success probability estimate. Moreover, this process should be performed for different values of the hyperparameters of SA to obtain an optimized value of the success probability. This becomes more inefficient as the size of the problems increases. Instead, one can use the PA algorithm to estimate and study two similar quantities, namely the \textit{mean square family size}~\cite{amey2018analysis}, $\rho_t$, and the \textit{entropic family size}~\cite{wang2015population}, $\rho_s$, that can be used as criteria for the correct thermalization of the algorithm, as well as for characterizing the complexity of different problems through the study of the free-energy landscape ruggedness. These quantities can be obtained from a single PA run for each problem instance, and have already been used for studying problem hardness since they correlate with other metrics such as the free energy systematic error~\cite{amey2018analysis}, the time-to-solution~\cite{perera2020computational}, and the integrated autocorrelation time in Markov Chain Monte Carlo~\cite{wang2015population}. Both of these quantities are obtained from the fraction of the final population that descends from the $i$-th replica of the initial population, $n_i$. Using this definition, the mean square family size is given by:
\begin{equation}
    \rho_t = \lim_{R\rightarrow \infty} R\cdot \sum^R_{i=1} n_i^2.
\end{equation}
Similarly, the entropic family size is defined as:
\begin{equation}
    \rho_s = \lim_{R\rightarrow \infty} R\cdot \exp{\left[\sum^R_{i=1} n_i\log(n_i)\right]}.
\end{equation}
Large values of $\rho_t$ and $\rho_s$ imply a low survival probability of the initial families and, thus, a more rugged energy landscape. Since both of the quantities are related and provide similar results~\cite{wang2015population}, we will focus on the study of $\rho_t$. Another quantity of interest can be obtained as follows: Consider that we obtain an uncorrelated set of replicas with their corresponding energies. The variance of the mean energy, $\sigma^2(\overline{E})$, given the energy variance, $\sigma^2(E)$, decreases with $R$ as:
\begin{equation}
    \sigma^2(\overline{E})=\frac{\sigma^2(E)}{R}.
\end{equation}
However, in practice, the MCMC steps might not completely remove the correlations introduced by the resampling step. For this case, we can model the effect of these correlations by considering an \textit{effective population size}, $R_{\text{eff}}$, defined as~\cite{weigel2021understanding}:
\begin{equation}
    R_{\text{eff}}\equiv \frac{\sigma^2(E)}{\sigma^2(\overline{E})}.
\end{equation}
Therefore, by estimating the values of these variances, one can estimate the corresponding $R_{\text{eff}}$. Given the replicas at some temperature, the value $\sigma^2(E)$ is straightforward to estimate, and $\sigma^2(\overline{E})$ can be obtained through means of jackknife estimation~\cite{weigel2021understanding, efron1994introduction}. In the following section, we will use these quantities to measure the diluted SK model hardness as a function of the parameter $p$.

\section{Problem hardness}
\label{sec:problem_hardness}

We perform simulations applying the PA algorithm to problem instances generated with different sizes $N$ and values of $p$. We set the inverse temperature to $\beta=5$ as the final inverse temperature for these simulations. We also set the remaining hyperparameters as $N_S=10$, $N_T=100$, and $R=10^5$. For these values of the hyperparameters and problem instances under consideration, we make sure that $R > 100\rho_t$ as a condition for thermal equilibration, which ensures that the values of interest found are close to their true value.

\begin{figure}[ht]
\centering
\includegraphics[width=1.0\columnwidth]{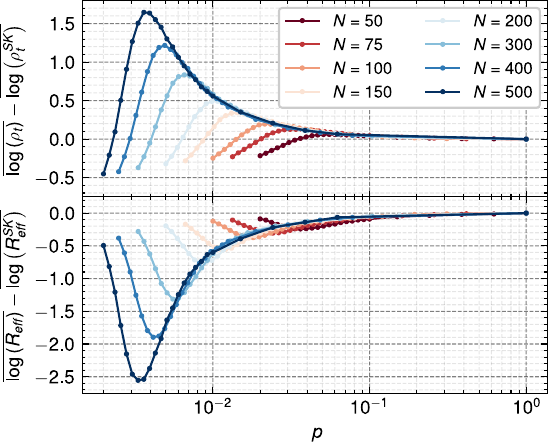}
\caption{Mean value of the logarithm of $\rho_t$ and $R_{\text{eff}}$ as a function of $p$ obtained for different problem sizes and values of $p$ normalized with respect to the corresponding mean value for that size at the SK limit. Error bars of these mean values are omitted as they are smaller than the marker sizes.}
\label{fig:rho_t_eff_diluted_p}
\end{figure}

\begin{figure}[ht]
\centering
\includegraphics[width=1.0\columnwidth]{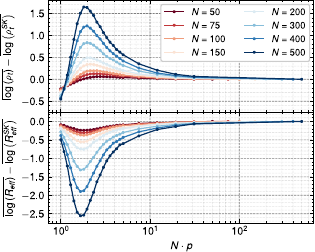}
\caption{Mean value of the logarithm of $\rho_t$ and $R_{\text{eff}}$ as a function of $Np$ obtained for different problem sizes and values of $p$ normalized to the corresponding mean value for each size at the SK limit. Error bars of these mean values are omitted as they are smaller than the marker sizes.}
\label{fig:rho_t_eff_diluted}
\end{figure}

\begin{figure}[t]
\centering
\includegraphics[width=0.95\columnwidth]{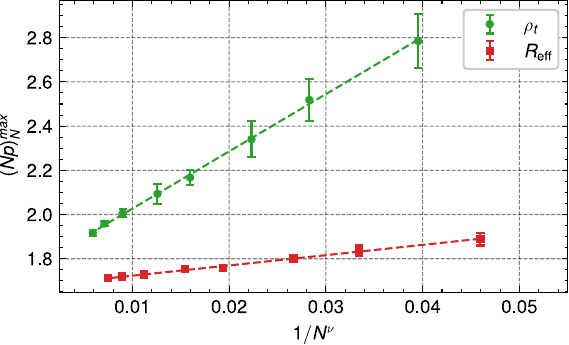}
\caption{Results of the fit from Eq.~\eqref{eq:asymptotic_peak} (discontinuous lines) to the values of $Np$ where the maxima of $\rho_t$ and $R_{\text{eff}}$ occurs for the results in Fig.~\ref{fig:rho_t_eff_diluted}. The error bars are obtained through bootstrap sampling.}
\label{fig:pth_fit}
\end{figure}

We show the mean values of $\log(\rho_t)$ and $\log(R_{\text{eff}})$ obtained from these simulations as a function of $p$ and for different problem sizes in Fig.~\ref{fig:rho_t_eff_diluted_p}. Each data point has been obtained by simulating $6000$ problem instances and it is normalized to the corresponding mean value of $\log(\rho_t)$ and $\log(R_{\text{eff}})$ for that size at the SK limit, i.e., $p=1$. We can see from these results that the relative values of $\rho_t$ and $R_{\text{eff}}$ with respect to the SK values increase as $p$ decreases similarly for all the problem sizes. However, these values show a peak at different values of $p$ for each problem size. In Fig.~\ref{fig:rho_t_eff_diluted} we show the same results as a function of $Np$. From this, we can see that the maximum values of $\rho_t$ and $R_{\text{eff}}$ approach a value close to $Np\approx 2$ as the problem size increases. We observe from these results that this model presents an easy-hard-easy transition in the problem family hardness as a function of $p$. Since this behaviour is similar to the one found in Ref.~\cite{perera2020computational}, we perform a similar study to better understand this behaviour. We obtain a rough estimate of the asymptotic value of the hardness peak as $N\rightarrow \infty$ from the results shown in Fig.~\ref{fig:rho_t_eff_diluted}. To do this, we fit the data points that are close to the peak to a third-order polynomial to estimate the peak positions for that problem size, $(Np)^{\text{max}}_{N}$, for each size considered. We find that the finite-size scaling of the hardness peak is well described by a function of the form:
\begin{equation}
\label{eq:asymptotic_peak}
    (Np)^{\text{max}}_{N} = (Np)^{\text{max}}_{N\rightarrow\infty} + A/N^\nu,
\end{equation}
from which we can estimate $(Np)^{\text{max}}_{N\rightarrow\infty}$. 
The peak values for each problem size and the corresponding fit are shown in Fig.~\ref{fig:pth_fit}. 
The power-law dependence of the finite-size scaling of the hardness peak position could be an indication of critical behaviour. 
However, a conclusive interpretation of the exponent $\nu$ in terms of critical exponents is beyond the scope of this work.
We present, for completeness, the fitted $A$, and $\nu$ values: 
(i) Hardness peak for $\rho_t$:  $(Np)^{\text{max}}_{N\rightarrow\infty, t}=1.77\pm 0.03$, $A_t=25.9\pm 4.4$, and $\nu_t=0.83\pm 0.05$; and (ii) Hardness peak for  $R_{\text{eff}}$: $(Np)^{\text{max}}_{N\rightarrow\infty, R} = 1.68\pm 0.01$, $A_R=4.7\pm 1.6$, and $\nu_R=0.79\pm 0.1$. 
A more accurate estimation of those values could be obtained by performing more simulations and increasing the size of the problem instances.

\begin{figure}[t]
\centering
\includegraphics[width=1.0\columnwidth]{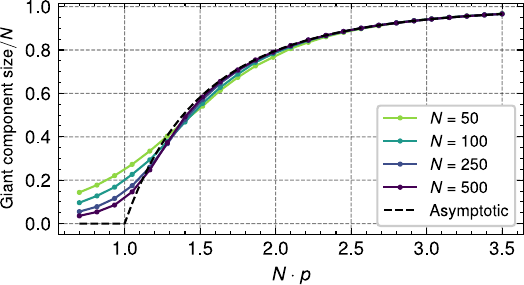}
\caption{Size of the giant component in Erd\H{o}s-R{\'e}nyi graphs for different values of $N$ and $Np$ obtained by calculating the average of 5000 instances for each data point. We also show the asymptotic value for $N\rightarrow \infty$. Error bars of these averaged values are omitted as they are smaller than the marker sizes.}
\label{fig:giant_component_study}
\end{figure}

To understand this easy-hard-easy transition of the diluted SK model, we can consider the properties of the underlying Erd\H{o}s-R{\'e}nyi graphs. It is known that, for these graphs, there is a transition that causes the biggest cluster of the graph (also called the \textit{giant component}) to be of size $O(N)$ for values $Np>1$. The average size of the giant component for different values of $N$ and $Np$ obtained from the simulation of 5000 instances are shown in Fig.~\ref{fig:giant_component_study}. Moreover, we also show the asymptotic value that is expected as $N\rightarrow \infty$~\cite{bollobas1998random, frieze2015introduction}. Therefore, we can expect two effects as a consequence of increasing the value of $Np$: First, smaller independent clusters start to connect, resulting in bigger clusters. This is important since one would expect that a problem instance composed of a set of small clusters would be easier to solve than the case in which all these spins are interacting, forming a giant cluster. Secondly, these clusters will show higher interconnectivity between their spins as $Np$ increases. Consequently, there will be fewer cases such as the one explained in Fig.~\ref{fig:diluted_example}, which are expected to increase the hardness of a problem instance. Therefore, higher connectivity can be expected to decrease the problem hardness. It is at intermediate connectivities that we expect weakly coupled clusters leading to the situation in Fig.~\ref{fig:diluted_example} and harder problems.

\begin{figure}[t]
\centering
\includegraphics[width=1.0\columnwidth]{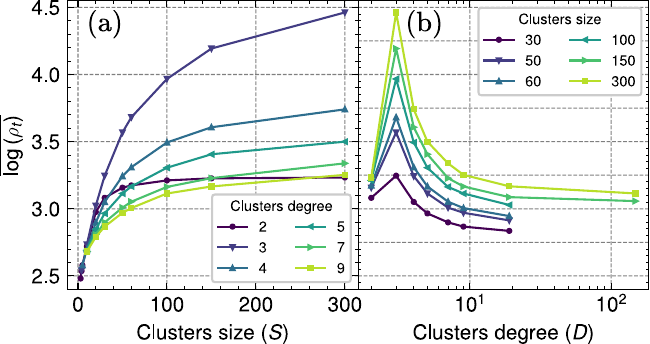}
\caption{Study of the problem hardness of clusterized regular graphs for problems of size $N=300$. (a) Behaviour of $\rho_t$ as the cluster sizes increase for different cluster degrees. The result of moving from small independent clusters to a big cluster is an increase in problem hardness. (b) Behaviour of $\rho_t$ as the cluster degree increases for different cluster sizes. As the connectivity of the clusters increases, the problem hardness is considerably reduced. Error bars of these mean values are omitted as they are smaller than the marker sizes.}
\label{fig:cluster_degree_study}
\end{figure}

To test these claims, we simulate the effects of the clusterization and the connectivity of each cluster on problem hardness. We consider problem instances with a total of $N=300$ spins for which we generate spin clusters of the same size $S$ but disconnected from each other. We also consider different connectivities by considering that each cluster is a regular graph of degree $D$. The weights of each interaction are then generated in the same way as explained for the diluted SK model. As an example, $S=100$ means that we consider three clusters, each of them consisting of $100$ spins with spins in each cluster having no interactions with spins in a different cluster. Inside each cluster the spins interact with other $D$ spin within the same cluster. We build these models by creating each cluster with connectivity $D$ and then we combine the clusters into the same model, resulting in a block diagonal interaction matrix $J$. We call these models \textit{clusterized regular graphs}. In the following, we only show results for $\rho_t$, since both $\rho_t$ and $R_{\text{eff}}$ yield similar results. The results are shown in Fig.~\ref{fig:cluster_degree_study}, where we obtain the values of $\log(\rho_t)$ for each instance and average over $1000$ instances for each set of parameters. In Fig.~\ref{fig:cluster_degree_study} (a) we study the behaviour of $\rho_t$ as the clusters size, S, increases, for different values of D. We observe the expected behaviour where, independent of the degree (or connectivity) of the clusters, the hardness of the problem instances increases as the smaller clusters are combined into a set of bigger clusters. This mechanism explains the easy-hard transition found at the left of the hardness peak in Fig.~\ref{fig:rho_t_eff_diluted}, where the values of $Np$ are low resulting in small independent clusters. Similarly, in Fig.~\ref{fig:cluster_degree_study} (b), we study how the connectivity of each cluster affects the problem hardness for different cluster sizes, including the case where all the spins are part of the same cluster, S=300. From this, we observe the same behaviour for all of these cases, where a peak of hardness appears for $D=3$, but then the hardness drops considerably as the connectivity increases. This explains the hard-easy transition that appears at the right of the hardness peak in Fig.~\ref{fig:rho_t_eff_diluted}. Taking both effects into consideration explains the easy-hard-easy behaviour found in Fig.~\ref{fig:rho_t_eff_diluted} for the diluted SK model.

\begin{figure}[t]
\centering
\includegraphics[width=1.0\columnwidth]{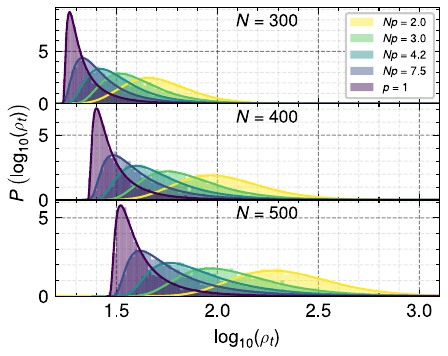}
\caption{Probability distribution of $\log_{10}(\rho_t)$ for different problem sizes and values of $p$. The continuous lines represent the corresponding fits to the log-inverse Gaussian distribution in Eq.~\eqref{eq:loginverse}.}
\label{fig:rho_t_histogram}
\end{figure}

After studying the easy-hard-easy transition of the diluted SK model, we now move on to study the distributions of $\log(\rho_t)$ for problem instances with different values of the problem size and $p$. The computational hardness of Ising spin glasses has a broad distribution with respect to different disorder realizations~\cite{wang2015population, yucesoy2013correlations}. Specifically, $\rho_t$ was found to be approximately distributed as a log-inverse Gaussian distribution when considering 3D Edward-Anderson spin glasses in Ref.~\cite{amey2018analysis}. This distribution is given by:
\begin{equation}
\label{eq:loginverse}
    P(x; \mu, \lambda, l) = \sqrt{\frac{\lambda}{2\pi(x-l)^3}} \exp\left[\frac{-\lambda(x-\mu-l)^2}{2\mu^2(x-l)}\right],
\end{equation}
where $x=\log_{10}(\rho_t)$, $l>0$ is a parameter that shifts the distribution, and $\lambda > 0$ is a shape parameter. Using the results of the simulations, we approximate the probability distribution of $\log(\rho_t)$ by considering histograms. We fit these histograms to the distribution in Eq.~\eqref{eq:loginverse} and show the results in Fig.~\ref{fig:rho_t_histogram}. Additionally, the parameters obtained from the fit for $p=1$ and $Np=2$ are shown in Table~\ref{table:parameters}. From these results, we can confirm that, for a fixed problem size, the problem hardness increases as the value of $Np$ decreases, until reaching a hardness peak. Moreover, the distribution not only moves to higher values of $\rho_t$ but also becomes wider, implying a broader distribution of the problem instance hardness.

\begin{table}[h!]
\centering
\begin{tabular}{p{1cm} p{2.25cm} p{2.25cm} p{2.25cm}}
 \hline
 \hline\\[-2ex]
  & \hspace{0.5em}$N=300$ & \hspace{0.5em}$N=400$ & \hspace{0.5em}$N=500$ \\ [0.5ex] 
 \hline\\[-2ex]
 $l$ & (0.93, 1.23) & (0.94, 1.35) & (0.99, 1.45) \\ 
 $\mu$ & (0.78, 0.12) & (1.09, 0.15) & (1.36, 0.17) \\
 $\lambda$ & (16.3, 0.13) & (27.0, 0.17) & (38.7, 0.23) \\[0.5ex]
 \hline
 \hline
\end{tabular}
\caption{Parameters obtained from the fits to Eq.~\eqref{eq:loginverse} and shown in Fig.~\ref{fig:rho_t_histogram} for different sizes and for $p=1$ (left values) and $Np=2$ (right values).}
\label{table:parameters}
\end{table}

\section{Comparison of the combinatorial optimization efficiency}
\label{sec:PAvsSA}

\subsection{Comparison between SA and PA}

We compare the efficiency in finding ground states for both the SA and PA algorithms. To do this, we consider problems with fixed size $N=200$ for different values of $Np=1.3, 2, 5, 200$. We find the exact ground state configuration and energy for each problem instance with $Np=1.3, 2, 5$ by using the Gurobi Optimizer~\cite{gurobi}. However, for $Np=200$ the solver becomes prohibitively slow due to the high connectivity. For these cases we find the ground state for each problem instance by performing PA with $R=10^5$, $N_T=100$, $N_S=20$, and target inverse temperature $\beta_f=5$. Similarly to the previous section, we require a large number of independent families at the final temperature to be confident about the ground state found by ensuring that the resulting values of $\rho_t$ satisfy $R\ge 100\rho_t$. Of course, population annealing is a heuristic algorithm and it is not guaranteed that the found states are the real ground states. However, in the following study, we have checked that neither the simulation annealing nor the population annealing that we used to benchmark the algorithms managed to find a configuration with a lower energy. Moreover, we note that with these hyperparameters we find the same minimum energies as Gurobi for the other problem instances with $Np=1.3, 2, 5$ (which are close to the complexity peak found in the previous section), so we expect the found solutions for the easier instances with $Np=200$ to be the ground state energies. Therefore, we will use these found energies as target energies for the combinatorial optimizations for the problems with $Np=200$.

\begin{figure}[t]
\centering
\includegraphics[width=1.0\columnwidth]{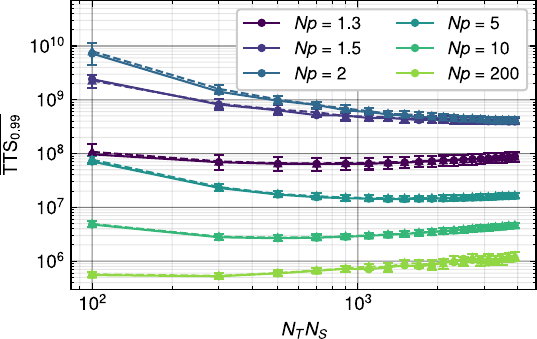}
\caption{Mean value of the TTS obtained over $10^3$ problem instances of size $N=200$. Two different values of $N_T$ have been simulated: $N_T=50$ (triangles with discontinuous lines), and $N_T=100$ (circles with continuous lines). Both values of $N_T$ result in similar efficiencies, which depend on $N_T N_S$. The error bars represent the standard deviation of the obtained mean values.}
\label{fig:TTS_scan_SA}
\end{figure}

\begin{figure*}[t]
\centering
\includegraphics[width=\textwidth]{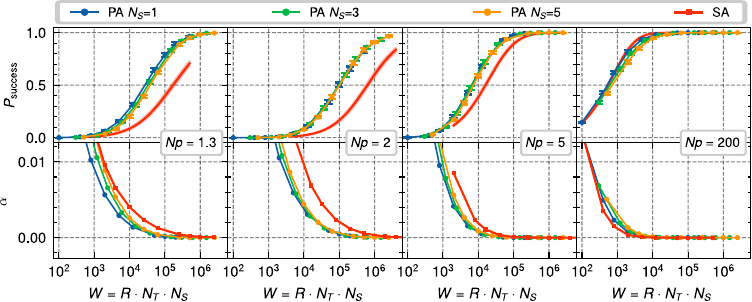}
\caption{Probability of finding the ground state, $P_\text{success}$ (top row), and approximation ratio, $\alpha$ (bottom row), averaged over $10^3$ problem instances of size $N=200$ for increasing computational resources for PA (with different values of $N_S$) and SA with optimized hyperparameters. PA shows a considerable advantage over SA for both metrics as the value $Np$ decreases. The error bars (for PA results) and the shaded area (for SA results) represent the $95\%$ confidence interval for the obtained mean values. The bottom figures do not include these intervals since they are smaller than the symbols.}
\label{fig:PAvsSA_scan}
\end{figure*}

Having found these minimum energy configurations for each problem instance, we perform an optimization of the SA parameters. We do this to guarantee that any observed advantage of PA is not due to suboptimal parameter choices for SA, but rather reflects a genuine improvement in efficiency. We do this by fixing the number of temperature steps $N_T$ and finding the mean TTS value as a function of $N_S$. Small values of $N_S$ do not guarantee the correct convergence of the Monte Carlo algorithm to a Boltzmann distribution, while too large values lead to an unnecessary long TTS. Each TTS value is found by calculating the value $g(1)$ for each problem and then using Eq.~\eqref{eq:gR} to extrapolate and calculate the mean TTS. The results are shown in Fig.~\ref{fig:TTS_scan_SA}. We do this for $N_T=50, 100$ and observe a similar behaviour for both cases. In the following, we will use $N_T=100$ for all the simulations of SA and PA. We observe a considerable increase in the TTS as the problems move from the SK cases to the most diluted ones. We note that optimizing SA for sizes higher than $N=200$ would require a considerable amount of resources for values of $Np$ close to the hardness peak.

Using these results, we obtain the optimal hyperparameters for the SA algorithm for different values of $Np$. Then, we find the probability of success of SA as a function of the total computational work $W=R N_T N_S$. We also calculate the success probability of PA for $N_S=1,5,10$. Since the replicas are not independent for the PA case, we cannot use Eq.~\eqref{eq:gR}. Instead, we perform simulations of PA for increasing values of $R$ several times to obtain estimates of the probability of success. We note that, while we do not do it in this work, one could consider using weighted averages~\cite{machta2010population, ebert2022weighted} over these repeated experiments to reduce the errors in the resulting estimates. The results of these simulations are shown in Fig.~\ref{fig:PAvsSA_scan}. As we can see, the behaviour of the relative efficiency between SA and PA depends considerably on the value of $Np$. For the $p=1$ cases, PA seems to behave worse than SA, but this is a consequence of using a higher value of $N_S$ than required. This effect is reduced when considering $N_S=1$, for which SA and PA behave similarly. This efficiency might be further improved but, in our implementation, we do not consider fewer spin-flips than a sweep per temperature step. More interestingly, PA shows an increasing improvement over SA as the value of $Np$ decreases. For the value $Np=2$ the PA algorithm shows a clear efficiency improvement compared to SA. Both the TTS and the relative improvement of PA over SA are decreased again for the value of $Np=1.3$, which is away from the hardness peak.

Finally, we perform a similar study considering a disorder-averaged approximation ratio to the ground state energy, given by:
\begin{equation}
    \alpha = 1 - \overline{(E_\text{min}/E_0)},
\end{equation}
where $E_0$ is the minimum energy of a problem instance and $E_\text{min}$ is the minimum energy found by SA or PA. The results of this study are shown in Fig.~\ref{fig:PAvsSA_scan}. From this, we can observe a similar behaviour as for the probability of finding the ground state. For the case with $Np=2$,  the PA algorithm shows a considerable improvement in accuracy compared to SA, which becomes less noticeable as $Np$ moves away from the hardness peak.

\subsection{Comparison with adaptive temperature schedule}

\begin{figure*}[t]
\centering
\includegraphics[width=\textwidth]{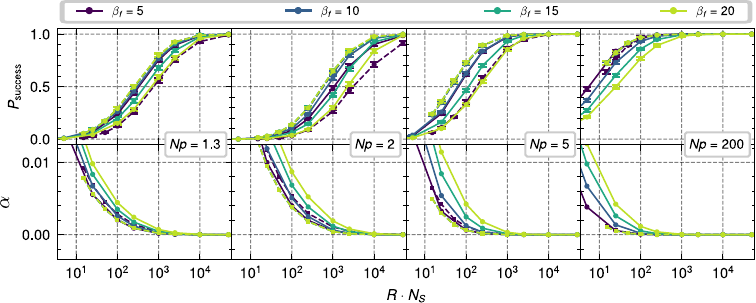}
\caption{Probability of finding the ground state, $P_\text{success}$ (top row), and approximation ratio, $\alpha$ (bottom row), using PA with a linear inverse temperature schedule (continuous lines) and an adaptive one (dashed lines), averaged over $10^3$ problem instances of size $N=200$ for increasing computational resources and for different final values of the inverse temperature. Curves representing the PA adaptive case are overlapped for $\beta = 10, 15, 20$. The error bars represent the $95\%$ confidence interval for the obtained mean values. The bottom figures do not include these intervals since they are smaller than the symbols.}
\label{fig:PAadaptive_scan_GS}
\end{figure*}

We also compare the relative efficiencies of PA when using a linear inverse temperature schedule and an adaptive one. We use the adaptive schedule described in Sec.~\ref{sec:population_annealing} with a culling fraction of $\epsilon=0.085$ and increase the inverse temperature until it reaches a target value of $\beta_f$. We set a minimum number of replicas, $R_\text{min}=15$, for the adaptive simulations since they require estimating the energy standard deviation at each temperature step. We do this for $\beta_f=5,10,15,20$. This approach results in a total of $N_T\approx 100$ temperature steps (with some deviation for each problem instance) for each value of $Np$ considered. This resulting value of $N_T$ obtained for each application of the adaptive PA is used for the linear schedule case to ensure that both cases use the same computational resources for each problem instance. The results of these simulations are shown in Fig.~\ref{fig:PAadaptive_scan_GS}. From this we observe that the efficiency depends on the value of $\beta_f$ for the linear case, with $\beta_f=10$ being the best for the cases close to the hardness peak, while $\beta_f=5$ is better as the problem instances get easier. This indicates that the linear schedule requires a calibration of the target temperature to achieve optimal performance, with the efficiency being worse for values of $\beta_f$ below or above this optimal value. As for the adaptive case, we observe that it either reaches a better performance than the linear case for the cases close to the hardness peak, or a similar performance for the easier cases. This is not the case for $\beta_f=5$, which is apparently a low value. Moreover, the performance of the adaptive schedule seems to be the same as long as $\beta_f$ is high enough to sample ground state configurations. This makes the adaptive schedule more robust against miscalibrations of the final temperature than the linear schedule.

Finally, we also study the approximation ratios when using both annealing schedules. We follow a similar approach as before, with the results being shown in Fig.~\ref{fig:PAadaptive_scan_GS}. From this, we can see similar advantages of the adaptive schedule with respect to the linear one as the ones observed for the probability of finding the ground state. The adaptive schedule is robust in the final temperature chosen while the linear schedule depends on a final temperature that must be calibrated to achieve an optimal efficiency. Furthermore, the adaptive schedule achieves smaller approximation ratios than those obtained from the linear schedules considered. Therefore, from the study of both the probability of finding the ground state and the approximation ratio, we conclude that using an adaptive schedule provides improved efficiency and less dependency on hyperparameters than the linear schedule, without any observed downsides.

\section{Conclusion}
\label{sec:conclusions}

In this work, we have studied the problem complexity of the diluted Ising model under the heuristic Population Annealing (PA) algorithm, and we have used this model to benchmark the relative efficiency of PA compared to Simulated Annealing (SA). Using the mean square family size, $\rho_t$, and the effective population size, $R_{\text{eff}}$, we managed to quantify the free-energy landscape ruggedness as a function of the problem size, $N$, and a parameter, $p$, which controls how much the problems are diluted. From this, we find an easy-hard-easy transition on the hardness of the problems as a function of $Np$. By studying the effect that the cluster sizes and the connectivity of each cluster have on problem complexity, we conclude that the asymptotic peak of the hardness transition is a consequence of the clusterization and connectivity behaviour of the underlying Erd\H{o}s-R{\'e}nyi graphs. Moreover, we also study the probability distribution of $\rho_t$ by fitting the resulting values to a log-inverse Gaussian distribution, as in Ref.~\cite{amey2018analysis}. The fitted distributions are in perfect agreement with the data and, by knowing that $\rho_t$ follows these distributions, one can estimate if the computational resources considered to study some instances of the same problem family are enough to solve new instances of the same family. Additionally, these fits show that the increase in hardness as $Np$ approaches the peak is not only related to higher mean values of $\rho_t$ but also to a spreading of the distribution, meaning that the harder problem instances of a family given by a value $Np$ require considerable more computational resources than the corresponding easier instances. As for the Sherrington-Kirkpatrick (SK) limit, $p=1$, the distribution has both a smaller mean value and variance, meaning that these problems are easier and each instance requires a similar amount of resources to thermalize.

Having found the hardness behaviour of the diluted SK model, we studied the relative efficiency between the SA and PA algorithms to expand the results found in Ref.~\cite{wang2015comparing}. To do this, we follow a similar procedure, where we fully calibrate the hyperparameters of SA for a problem size $N=200$ and different values of $p$. Using this optimally calibrated SA, we compare it with PA by studying both the probability of success and the approximation ratio as a function of the computational resources. For both of these quantities, we found that the PA algorithm outperformed the SA algorithm for values of $Np$ close to the hardness peak, with this efficiency improvement vanishing as the problems move closer to the SK limit. We assume that this behaviour results from the free-energy landscape becoming less rugged as the problem families reach the SK limit, making the resampling step from PA less important. However, we note that these results are obtained for a fixed and small value of $N$ that allowed us to study the relative efficiency for values of $Np$ close to the hardness peak, and a higher ruggedness of the free-energy landscape as $N$ increases could result again in an advantage of PA over SA in the SK limit.

Finally, we perform a similar efficiency study comparing two cases of PA: one using a linear inverse temperature schedule and another using an adaptive one. From this study, we find that the efficiency of the linear schedule is considerably dependent on the chosen final inverse temperature. However, the adaptive schedule seems independent of the final temperature as long as it is high enough to sample the ground state configuration, making this type of schedule more robust under miscalibrations of the final inverse temperature. Moreover, the adaptive schedule showed a slight efficiency improvement for problems close to the hardness peak compared to the best final inverse temperature found for PA with a linear schedule.

Our work shows that the PA algorithm is ideally suited for combinatorial optimization problems of varying connectivity. This is relevant for practical applications where PA has the additional advantage of providing us with an adaptive method for hyperparameter selection. Moreover, an additional feature of PA is that problem hardness can be estimated by measures like the entropic family size.

Manuscript data is available upon request from the authors.

\section{Acknowledgments}
This project has been supported by Spanish Projects No. PID2021-127968NB-I00 and No. PDC2022-133486-I00, funded by MCIN/AEI/10.13039/501100011033,  by the European Union “NextGenerationEU”/PRTR”1; and CSIC Interdisciplinary Thematic Platform (PTI) Quantum Technologies (PTI-QTEP). DP acknowledges support by the Ministry for Digital Transformation and of Civil Service of the Spanish Government through the QUANTUM
ENIA project call - Quantum Spain project, and by the European Union through
the Recovery, Transformation and Resilience Plan - NextGenerationEU within
the framework of the Digital Spain 2026 Agenda.
We acknowledge both the Scientific Computing Area (AIC), SGAI-CSIC, for their assistance while using the DRAGO Supercomputer and Centro de Supercomputación de Galicia (CESGA) who provided access to the supercomputer FinisTerrae for performing the simulations.


\bibliography{sample}

\end{document}